\documentstyle[epsf]{article}
\def\ru1{\rule[-0.21truecm]{0mm}{0.1truecm}}
\begin{document}

{\hfill DSF-99-35}
\begin{center}
{\Large \bf Decay Constants and Semileptonic Form Factors of }\\
\vspace{3truemm} {\Large \bf Pseudoscalar Mesons\footnote{Based on
talk given by the first author at the {\it QFTHEP '99} Workshop,
Moscow, May 27- June 2, 1999.}}
\\ \vspace{4truemm}
 M.A. Ivanov$^{a}$ and  P. Santorelli$^{b}$\\
 \vspace{2truemm}
  $^a$ {Bogoliubov
 Laboratory of Theoretical Physics, \\ Joint Institute for Nuclear
 Research, 141980 Dubna, Russia}\\
 $^b$ {Dipartimento di Scienze Fisiche, \\
 Universit\`a  ``Federico II" di Napoli, Napoli, Italy\\
 and INFN Sezione di Napoli}
 \end{center}

\begin{abstract}
\noindent
A relativistic constituent quark model is adopted to give an unified
description of the leptonic and semileptonic decays of pseudoscalar
mesons ($\pi$, $K$, $D$, $D_s$, $B$, $B_s$).
The calculated leptonic decay constants and form factors are found to be
in good agreement with available experimental data and the results of
other approaches. Eventually, the model is found to reproduce the scaling
behaviours of spin-flavor symmetry in the heavy-quark limit.
\end{abstract}

 \section{Introduction}
Semileptonic decays of pseudoscalar mesons allow to evaluate the
elements of the Cabibbo--Kobayashi--Maskawa (CKM) matrix, which are
fundamental parameters of the Standard Model. The decay $K\to\pi
e\nu$ provides the most accurate determination of $V_{us}$, the
semileptonic decays of D and B mesons, $D\to K(K^*) l\nu$, $B\to
D(D^*)l\nu$ and $B\to \pi(\rho)l\nu$, can be used to determine
$|V_{cs}|$, $|V_{cb}|$ and $|V_{ub}|$, respectively. This program
can be performed if the non-perturbative QCD effects, which are
parameterized by the form factors, are known. Up to now, these
form factors cannot be evaluated from first principles, thus models,
more or less connected with QCD, are usually considered for this
purpose. Here we discuss a relativistic quark model \cite{noi},
previously  used to study the baryon form factors \cite{RCQM}.

This model is based on an effective Lagrangian describing the
coupling of mesons with their constituent quarks. The physical
processes are described by the one-loop quark diagrams and
meson-quark vertices related to the Bethe-Salpeter amplitudes. In principle,
the vertex functions and quark propagators should be
given by the Bethe-Salpeter and Dyson-Schwinger equations,
respectively. This kind of analysis is provided by the
Dyson-Schwinger Equation (DSE) studies \cite{DSE} and an unified
description of light and heavy meson observables was carried out
in \cite{DSEH1,DSEH2}. Here, instead, we use free propagators
for constituent quarks and consider a Gaussian vertex function
as Bethe-Salpeter confining function. The adjustable
parameters, the widths of Bethe-Salpeter amplitudes in momentum space,
and the constituent quark masses, are determined from the best fit
of available experimental data and some lattice simulations.
Our results are in good agreement with experimental data and other approaches.
We also reproduce the spin-flavor symmetry relations and scaling for
leptonic decay constants and semileptonic form factors in the heavy-quark
limit \cite{IsgurWise}.

An approach similar to the one presented here, can be found in
\cite{Gatto}. That  model is based on meson-quark interactions,
so the mesonic transition amplitudes are described by diagrams
with heavy mesons attached to quark loops. The free propagator
has been used for light quarks, while the quark propagator
obtained in the heavy-quark limit has been adopted for heavy
quarks.

\section{Our model}

Our starting point is the effective Lagrangian describing the
coupling between hadrons and quarks. The
\begin{equation}
\label{lag}
{\cal L}_{{\rm int}} (x)=g_H H(x) \int\!\! dx_1 \!\!\int\!\! dx_2
\Phi_H (x;x_1,x_2)
\bar q(x_1) \Gamma_H \lambda_H q(x_2)\,
\end{equation}
describes the transition of the meson $H$ ($\lambda_H$ is the corresponding
combination of Gell-Mann  matrices, $\Gamma_H$ are Dirac matrices) into its
constituents $q_1$ and $q_2$.
The function $\Phi_H$ is related to the scalar part of the Bethe--Salpeter amplitude.

The coupling constant $g_H$ is given by the derivative of
the meson mass operator $\widetilde\Pi_H$ by the
{\it compositeness condition} \cite{SW}:

\begin{equation}
\label{comp}
Z_H=1-\frac{3g^2_H}{4\pi^2}\widetilde\Pi^\prime_H(m^2_H)=0\,.
\end{equation}

It is worth noticing that, due to the absence of confinement,
the sum of constituent quark masses should be larger than the mass
of the corresponding meson otherwise, imaginary parts in
physical quantities appear. This allows us to consider
low-lying pseudoscalar mesons only.

Now, to give an example of the hadronic part of
invariant amplitudes, we will evaluate the form factor $f_+$,
which appears in the semileptonic decays of a pseudoscalar meson
into another one, $H \rightarrow H^{\prime} \ell \nu$. The
invariant amplitude can be written as:
\begin{equation}
A(H(p)\to H'(p')\, e\, \nu) = {G_F\, V_{q_1 q_2} \over \sqrt{2}}
\bigl [\bar \ell\, \gamma_{\mu}(1-\gamma_5)\, \nu \bigr ]
M^\mu_{HH^{\prime}}(p,p^\prime),
\end{equation}
where $G_F$ is the Fermi constant, $V_{q_1 q_2}$ is the corresponding
Cabibbo--Kobayashi--Maskawa matrix element,
and, in our model, $ M^\mu_{HH^{\prime}}(p,p^\prime)$ is given by:
\begin{eqnarray}
M^{\mu}_{HH^{\prime}}(p,p^\prime)&=&
{3\, g_H \, g_{H^{\prime}}\over 4\,\pi^2} \!\!\int\!\!{d^4k\over 4\pi^2i}
\phi_H(-k^2)\phi_{H^{\prime}}(-k^2)
{\rm tr}\biggl[\gamma^5 S_3(\not\! k)\gamma^5
S_2(\not\! k+\not\! p^\prime)\gamma^{\mu}(1-\gamma_5)
S_1(\not\! k+\not\! p) \biggr]
\label{e:Mmunu}\\
&&\nonumber\\
& = & {3\, g_H \, g_{H^{\prime}}\over 4\,\pi^2}
\biggl [ I_+(p^2,p^{\prime 2},q^2)(p+p^\prime)^\mu + I_-(p^2,p^{\prime 2},q^2)q^\mu
\biggr] \nonumber\\
&& \nonumber \\
& = &f_+(q^2)(p+p^\prime)^\mu + f_-(q^2)q^\mu \nonumber\,.
\end{eqnarray}
Here, $q\, = \, p-p^\prime$, $S_i(k) =1/(m_i -\not \! k)$ is the
propagator of the quark $i$ with mass $m_i$.

To evaluate the integral in Eq. (\ref{e:Mmunu}) we have to
calculate the following integrals
(${\cal F}(-k^2)=\phi_H(-k^2)\cdot\phi_{H^\prime}(-k^2)$):
\begin{equation}\label{int1}
J^{(0,\mu,\mu\nu,\mu\nu\delta)}\, \equiv \,\int \frac{d^4k}{\pi^2i}
\frac{(1,k^\mu,k^\mu k^\nu,k^\mu k^\nu k^\delta){\cal F}(-k^2)}
{[m_1^2-(k+p)^2][m_2^2-(k+p^\prime)^2][m_3^2-k^2]}\,.
\end{equation}
They can be evaluated using the standard Feynman
$\alpha$-representation and the integral Cauchy
representation for the function ${\cal F}$:
\begin{eqnarray}
J^{0}&=&
\int\limits_0^\infty dt \biggl(\frac{t}{1+t}\biggr)^2
\int\! d^3\alpha\,\delta\biggl(1-\sum\limits_{i=1}^3\alpha_i\biggr)
\biggl(-\frac{d {\cal F}(z_I)}{d z_I}\biggr)
\label{scalar}\\
&&\nonumber\\
&&\nonumber\\
J^{\mu}&=&
-\int\limits_0^\infty dt \biggl(\frac{t}{1+t}\biggr)^3
\int\! d^3\alpha\,\delta\biggl(1-\sum\limits_{i=1}^3\alpha_i\biggr)
P^\mu
\biggl(-\frac{d {\cal F}(z_I)}{d z_I}\biggr)
\label{vector}\\
&&\nonumber\\
&&\nonumber\\
J^{\mu\nu}&=&
\int\limits_0^\infty dt \biggl(\frac{t}{1+t}\biggr)^2
\!\!\int\! d^3\alpha\, \delta\biggl(1-\sum\limits_{i=1}^3\alpha_i\biggr)
\biggl\{
-\frac{1}{2}g^{\mu\nu}\frac{1}{1+t}{\cal F}(z_I)
-P^\mu P^\nu \biggl(\frac{t}{1+t}\biggr)^2
\biggl(-\frac{d {\cal F}(z_I)}{d z_I}\biggr)
\biggr\}
\label{tensor2}\\
\nonumber\\
&&\nonumber\\
J^{\mu\nu\delta}&=&
\int\limits_0^\infty dt \biggl(\frac{t}{1+t}\biggr)^2
\int\! d^3\alpha\, \delta\biggl(1-\sum\limits_{i=1}^3\alpha_i\biggr)
\biggl\{
\frac{1}{2}\biggl[g^{\mu\nu}P^\delta+
                  g^{\mu\delta}P^\nu+
                  g^{\nu\delta}P^\mu \biggr]
\frac{t}{(1+t)^2}{\cal F}(z_I) \, + \nonumber\label{tensor3}\\
&&\nonumber\\ &&\hspace{6.5truecm}P^\mu P^\nu P^\delta
\biggl(\frac{t}{1+t}\biggr)^3 \biggl(-\frac{d {\cal F}(z_I)}{d
z_I}\biggr) \biggr\}
\end{eqnarray}
where $P=\alpha_1 p+\alpha_2 p^\prime$,
$D_3=\alpha_1\alpha_3 p^2+\alpha_2\alpha_3 p^{\prime 2}+\alpha_1\alpha_2 q^2 $,
and
\begin{equation}
z_I=t\left [\sum_{i=1}^3\alpha_i m^2_i-D_3\right
]-\,\frac{t}{(1+t)}\,P^2.
\end{equation}

Finally, we have the analytical expression for $I_+$
\begin{eqnarray}
I_+(p^2,p^{\prime 2},q^2)&=& \frac{1}{2}\int\limits_0^\infty dt
\biggl(\frac{t}{1+t}\biggr)^2 \int\! d^3\alpha\,
\delta\biggl(1-\sum\limits_{i=1}^3\alpha_i\biggr) \biggr\{ {\cal
F}(z_I)\frac{1}{1+t}\biggl[4-3(\alpha_1+\alpha_2)\frac{t}{1+t}\biggr]
 \nonumber \label{intfin}\\
 &&\nonumber\\
 && + \biggl(-\frac{d
 {\cal F}(z_I)}{d z_I}\biggr) \biggl[(m_1+m_2)m_3
 +\frac{t}{1+t}\biggl(-(\alpha_1+\alpha_2)(m_1m_3+m_2m_3-m_1m_2)
\nonumber\\
 &&\nonumber\\ && +\alpha_1 p^2+\alpha_2 p^{\prime
2}\biggr) -P^2 \biggl(\frac{t}{1+t}\biggr)^2
\biggl(2-(\alpha_1+\alpha_2)\frac{t}{1+t}\biggr) \biggr]
\biggr\}\,,
\end{eqnarray}
which can be used also to obtain the normalization constants $g_H$
and $g_{H^{\prime}}$ in Eq. (\ref{comp}). For example, $g_H$ is
given by
\begin{equation} \label{normfin}
g_H = \sqrt{\frac{4\pi^2}{3 I_+(p^2,p^2,0)}}
\end{equation}
where we put $m_1=m_2\equiv m$.

\section{Results and discussion}

Once obtained the analytical expressions for the invariant
amplitudes, a comparison with the experimental data is in order.
For this purpose we have to specify the analytical form of
$\phi_H(-k^2)$ and the constituent masses appearing in the
expressions. In particular, we choose the Gaussian form
$\phi(-k^2)=\exp\{k^2/\Lambda_H^2\}$ in Minkowski
space, where the magnitude  of $\Lambda_H$ characterizes the size of the
BS-amplitude and is an adjustable parameter in our approach. Thus, to
describe processes involving $\pi$, $K$, $D$, $D_s$, $B$, and $B_s$  mesons,
we have six $\Lambda$-parameters plus four quark masses
($m_q = m_u=m_d$, $m_s$, $m_c$, and $m_b$), all of them are fixed
{\it via} the least-squares fit to the observables measured experimentally or
taken from lattice simulations (as is reported in Table~\ref{t1}).

\begin{table}[h]
\caption{Calculated values of a range of observables
($g_{\pi\gamma\gamma}$ in GeV$^{-1}$, leptonic decay constants in GeV,
form factors and ratios are dimensionless). The ``Obs." are extracted from
Refs.~\protect \cite{PDG,CLEO-BD,CLEO-Bpi,Flynn,Wittig,Debbio,MILC}.
The quantities used in fitting free parameters are marked by ``$\ast$".
\label{t1}}
\begin{center}
\begin{tabular}{clll|clll}
  &      & Obs.  & Calc. & & & Obs.  & Calc. \\
\hline
\ru1
$\ast$ & $g_{\pi\gamma\gamma}$   & 0.274                  & 0.242   &
       & $f_+^{K\pi}(0)$         & 0.98                   & 0.98  \\
\ru1
$\ast$ & $f_\pi$                 & 0.131                  & 0.131   &
$\ast$ & $f_+^{DK}(0)$           & 0.74 $\pm$ 0.03        & 0.74 \\
\ru1
$\ast$ & $f_K$                   & 0.160                  & 0.160   &
       & $f_+^{BD}(0)$           &                        & 0.73    \\
\ru1
$\ast$ & $f_D$                   & 0.191$^{+19}_{-28}$    & 0.191   &
       & $f_+^{B\pi}(0)$         & 0.27 $\pm$ 0.11        & 0.51    \\
\ru1
$\ast$ & $\frac{f_{D_s}}{f_D}$   & 1.08(8)                & 1.08    &
  & ${\rm Br}(K\to\pi l\nu)$&   $(4.82\pm 0.06)\cdot 10^{-2}$
                            &   $4.4\cdot 10^{-2}$            \\
\ru1
  & $f_{D_s}$               & 0.206$^{+18}_{-28}$     & 0.206   &
  & ${\rm Br}(D\to K l\nu)$ &   $(6.8\pm 0.8)\cdot 10^{-2}$
                            &   $ 8.1\cdot 10^{-2}$            \\
\ru1
$\ast$ & $f_B$                   & 0.172$^{+27}_{-31}$     & 0.172   &
       & ${\rm Br}(B\to D l\nu)$ &   $(2.00\pm 0.25)\cdot 10^{-2}$
                            &   $2.3\cdot 10^{-2}$            \\
\ru1
$\ast$ & $\frac{f_{B_s}}{f_B}$   & 1.14(8)                & 1.14    &
       & ${\rm Br}(B\to\pi l\nu)$&   $(1.8\pm 0.6)\cdot 10^{-4}$
                            &   $2.1\cdot 10^{-4}$            \\
\ru1
  & $f_{B_s}$               &                        & 0.196   &
  &                         &                        &         \\
\hline
\end{tabular}
\end{center}
\end{table}

The fitted values for the free parameter of our model are the
following:
\begin{eqnarray}
(\Lambda_\pi,\,\Lambda_K,\,\Lambda_D,\,\Lambda_{D_s},\,\Lambda_B,\,\Lambda_{B_s})
& = & (1.16,\,1.82,\,1.87,\,1.95,\,2.16,\,2.27)\,\,GeV \\
&& \nonumber\\
(m_q,\,m_s,\,m_c,\,m_b)& = & (0.235,\,0.333,\,1.67,\,5.06)\,\,GeV\,.
\end{eqnarray}

Note that the $\Lambda$ values are larger for
mesons with larger mass, {\it i.e.} $\Lambda_H \, <
\Lambda_{H^{\prime}}$ when $m_H\, < \, m_{H^{\prime}}$.
This correctly corresponds to the ordering law for the sizes of bound states.

The u(d)-quark mass and the parameter $\Lambda_{\pi}$ are almost
fixed from the rate $\pi\rightarrow\mu\nu$ and $\pi^0\rightarrow\gamma\gamma$
with an accuracy of a few percent. Moreover, the obtained value of
$m_u$ is less than the constituent-light-quark mass used in baryon
physics.

Let us now consider the $q^2$-behaviour of the form factors. Since
a numerical integration should be done (see, Eq. (\ref{intfin})),
 we do not have a simple analytical expression for them.
However, looking at the numerical results, the form
\begin{equation}\label{approx}
f_+^{HH'}(q^2)=\frac {f(0)} {\displaystyle 1-
b_0\left(\frac{q^2}{m_H^2}\right)-b_1\left(\frac{q^2}{m_H^2}\right)^2}
\end{equation}
is suitable for a good description of the $q^2$-behaviour, once
the parameters $b_0,~b_1$ and $f(0)$ are fixed.
Their values are collected in the following Table:
\begin{equation}
\label{coef}
\begin{array}{c|cccc}
            &\ \ K\to \pi\ \  & \ \ D\to K\ \  &\ \ B\to D\ \
&\ \ B\to\pi\ \  \\
\hline \ru1
     f(0)   & 0.98  & 0.64 & 0.73 & 0.51 \\ \ru1
     b_0    & 0.28  & 0.64 & 0.77 & 0.52 \\ \ru1
     b_1    & 0.057 & 0.20 & 0.19 & 0.38 \\
\end{array}\,
\end{equation}

It should be noted that a value for $f_+^{B\pi}(0)$ larger than
those obtained by QCD Sum Rules \cite{CSB}, and other approaches
\cite{noiBpi}. In any case, as one can see from the Table
\ref{t1}, the agreement between our predictions and experimental
data on semileptonic branching ratios is impressive.

As we have seen, our model gives an accurate and unified
description of the weak and radiative ($\pi^0\to \gamma\gamma$)
transitions involving pseudoscalar mesons. Moreover, as already
stated in the introduction, it is able to reproduce the scaling
behavior predicted by QCD in the heavy-quark limit. For more
details, see the original paper. Here we report the way to obtain
this limit in the expression for $f_+$. The heavy-quark limit
corresponds to consider $m_1\equiv M\to\infty$, $m_2\equiv
M'\to\infty$ and $p^2=(M+E)^2$, $p^{\prime 2}=(M'+E)^2$ with $E$
being a constant value independent of $M$ and $M'$. By replacing
in Eq.~(\ref{intfin}) the variables $\alpha_1$ with $\alpha_1/M$
and $\alpha_2$ with $\alpha_2/M'$, one obtains
\begin{eqnarray}
I_+ & \rightarrow & \frac{M+M'}{2MM'}\cdot \int\limits_0^\infty dt
\biggl(\frac{t}{1+t}\biggr)^2 \int\limits_0^1 d\alpha\alpha
\int\limits_0^1 d\tau \biggl({-\cal F'}(z)\biggr)
\biggl[m+\frac{\alpha t}{1+t}\biggr] \nonumber\\ &&\nonumber\\ &=&
\frac{M+M'}{2MM'}\cdot \frac{1}{2} \int\limits_0^1 \frac{d\tau}{W}
\int\limits_0^\infty du {\cal F}(\tilde
z)\frac{m+\sqrt{u/W}}{m^2+\tilde z}
\label{inthql}
\end{eqnarray}
where
\begin{equation}
 \tilde z=u-2E\sqrt{\frac{u}{W}},
 \hspace{1truecm}
 W=1+2\tau(1-\tau)(w-1),
 \hspace{1truecm} w=\frac{M^2+M^{\prime 2}- q^2}{2MM'}\,.
\end{equation}
Therefore, using the relation between $I_+$ and $f_+$ and the
normalization condition, the correct scaling relation is found:
\begin{equation}
f_+ \rightarrow \frac{M' \, + \, M}{2\sqrt{MM'}}\xi(w)
\hspace{1cm} \xi(w)\propto \int\limits_0^1 \frac{d\tau}{W}
\int\limits_0^\infty du \,\phi_H^2(\tilde z)\,
\frac{m+\sqrt{u/W}}{m^2+\tilde z}\,.
\end{equation}

In conclusion, we can see that the agreement with experimental
data and lattice results is very good, with the exception of the
value of $f_+^{bu}(0)$ which is found to be larger  than the
monopole extrapolation of a lattice simulation, QCD Sum Rules (cf.
\cite{CSB}) and some other quark models (see, for example,
\cite{BSW,noiBpi}). However, this result is consistent with the
value calculated from Refs.~\cite{DSEH2,infra} and a light-front
constituent quark model \cite{Simula}. Moreover, it allows us to
reproduce the experimental data on $B\to \pi l \nu$ decay.
\vspace{1truecm}

\noindent
{\bf Acknowledgements}\\
We thank G. Esposito for reading the manuscript.







\end{document}